\documentclass[12pt,aps,prd,preprint,tightenlines,
   showpacs,nofootinbib]{revtex4}
\newcommand{\PRE}[1]{{#1}} 

\usepackage{bm} \usepackage{epsfig}

\newcommand{\sigmaSI}{\sigma_{\text{SI}}}
\newcommand{\sigmabb}{\sigma_{b \bar{b}}}

\newcommand{\kev}{\text{keV}}

\newcommand{\gev}{\text{GeV}}
\newcommand{\tev}{\text{TeV}}
\newcommand{\pb}{\text{pb}}

\newcommand{\cm}{\text{cm}}

\newcommand{\s}{\text{s}}

\newcommand{\sr}{\text{sr}}

\newcommand{\etal}{{\em et al.}}
\newcommand{\eg}{{\em e.g.}}

\newcommand{\eqref}[1]{Eq.~(\ref{#1})}

\newcommand{\figref}[1]{Fig.~\ref{fig:#1}}

\newcommand{\be}{\begin{equation}}
\newcommand{\ee}{\end{equation}}
\newcommand{\ssection}[1]{{\em #1.}}

\begin{document}

\preprint{UCI-TR-2008-22, NSF-KITP-08-102}

\title{
\PRE{\vspace*{1.5in}}
Explaining the DAMA Signal with WIMPless Dark Matter
\PRE{\vspace*{0.3in}}
}

\author{Jonathan L.~Feng}
\affiliation{Department of Physics and Astronomy, University of
California, Irvine, CA 92697, USA
\PRE{\vspace*{.5in}}
}

\author{Jason Kumar%
\PRE{\vspace*{.2in}}
}
\affiliation{Department of Physics and Astronomy, University of
California, Irvine, CA 92697, USA
\PRE{\vspace*{.5in}}
}

\author{Louis E. Strigari%
\PRE{\vspace*{.2in}}
}
\affiliation{Department of Physics and Astronomy, University of
California, Irvine, CA 92697, USA
\PRE{\vspace*{.5in}}
}

\begin{abstract}
\PRE{\vspace*{.3in}} WIMPless dark matter provides a framework in
which dark matter particles with a wide range of masses naturally have
the correct thermal relic density.  We show that WIMPless dark matter
with mass around 2-10 GeV can explain the annual modulation observed by
the DAMA experiment without violating the constraints of other dark
matter searches.  This explanation implies distinctive and promising
signals for other direct detection experiments, GLAST, and the LHC.
\end{abstract}


\maketitle

\ssection{Introduction} Dark matter makes up 24\% of the energy
density of the Universe, but its identity is unknown.  At present all
incontrovertible evidence for dark matter is based on its
gravitational interactions.  As the requisite first step toward
identifying dark matter, diverse experiments worldwide are searching
for evidence for additional dark matter interactions, with several
tentative hints reported so far.

By far the most significance claimed for a non-gravitational signal is
the DAMA Collaboration's observation~\cite{Bernabei:2008yi} of annual
modulation~\cite{Drukier:1986tm} in recoil scattering off NaI(Tl)
detectors deep underground at the Gran Sasso National Laboratory.
When combined with previous results~\cite{Bernabei:2003za}, these
recent data yield an $8.2\sigma$ signal based on a total exposure of
0.82 ton-years.  The observed modulation has period $T = 0.998 \pm
0.003$ years and maximum at $t = 144 \pm 8$ days, both perfectly
consistent with the values $T = 1$ year and $t = 152$ days expected
for dark matter, given simple astrophysical assumptions.

Experimental aspects of the DAMA result have been the topic of lively
discussion, to which we have nothing to add.  {}From a theoretical
viewpoint, however, the DAMA result is also very interesting, because
it has not been easy to reconcile with other experimental constraints
or to explain with candidates that are motivated by considerations
other than the DAMA anomaly itself.  Of course, comparisons with other
experiments and theory are model-dependent, requiring additional
assumptions from both particle physics and astrophysics. At the same
time, it is likely that a definitive discovery of dark matter will
require confirmation by more than one experiment under the unifying
umbrella of a plausible theoretical framework.  Toward this end, we
here propose a dark matter explanation that has well-motivated
features and then determine other observable predictions that may be
used to exclude or favor the proposed explanation.

\ssection{DAMA Regions} The DAMA signal is consistent with the
scattering of dark matter particles $X$ through elastic,
spin-independent interactions.  The conventional region has mass and
$X$-nucleon cross section $(m_X, \sigmaSI) \sim (20 - 200~\gev,
10^{-5}~\pb)$~\cite{Belli:1999nz}.  This is now excluded, most
stringently by XENON10~\cite{Angle:2007uj} and CDMS
(Ge)~\cite{Ahmed:2008eu}, which require $\sigmaSI < 10^{-7}~\pb$
throughout this range of $m_X$.  Spin-dependent and other more general
couplings do not remedy the situation~\cite{Ullio:2000bv}.  Mirror
dark matter has been proposed as a solution~\cite{Foot:2008nw}, as has
inelastic scattering, where the dark matter particle is accompanied by
a companion particle that is roughly $\sim 100~\kev$
heavier~\cite{Smith:2001hy}.

\begin{figure}
\resizebox{3.25in}{!}{
\includegraphics{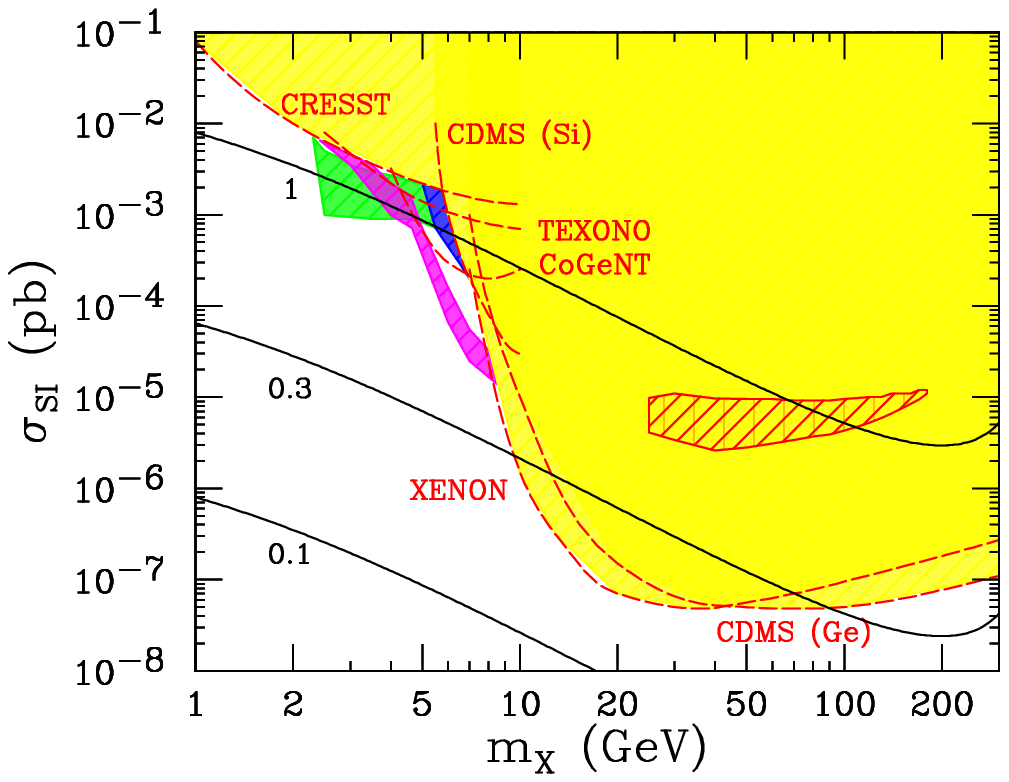}
}
\caption{Direct detection cross sections for spin-independent
$X$-nucleon scattering as a function of dark matter mass $m_X$.  The
solid curves are the predictions for WIMPless dark matter with
connector mass $m_Y = 400~\gev$ and the Yukawa couplings $\lambda_b$
indicated.  The light yellow shaded region is excluded by the
experimental results indicated (see text). The dark blue shaded region
is consistent with the DAMA signal at $3\sigma$, using 2-4 and 6-14
keVee bins; it may be extended to the medium green shaded region with
the inclusion of dark matter streams and 2-6 and 6-14 keVee
bins~\cite{Gondolo:2005hh}.  The medium-dark magenta shaded region is
DAMA-favored when channeling is included (but streams are
not)~\cite{Petriello:2008jj}.  The cross-hatched region is the
conventional DAMA-favored region~\cite{Belli:1999nz}, which is now
excluded by other experiments.
\label{fig:damadirect}
}
\end{figure}

Gondolo and Gelmini have noted, however, that an alternative region
with $(m_X, \sigmaSI) \sim (1-10~\gev, 10^{-3}~\pb)$ may explain the
DAMA results without violating other known
bounds~\cite{Gondolo:2005hh}. DAMA's relative sensitivity to this
region follows from its low energy threshold and the lightness of Na
nuclei.  This region is extended to lower masses and cross sections by
the effects of
channeling~\cite{Bernabei:2007hw,Petriello:2008jj,Bottino:2008mf} and
may also be broadened if dark matter streams exist in the solar
neighborhood~\cite{Gondolo:2005hh}, arising, for example, from the
destruction of Galactic satellites~\cite{Stiff:2001dq,Diemand:2008in}.
The allowed region is constrained by null results from
CRESST~\cite{Angloher:2002in}, CDMS (Si)~\cite{Akerib:2005kh},
TEXONO~\cite{Lin:2007ka} (see, however, Ref.~\cite{Avignone:2008xc}),
and CoGeNT~\cite{Aalseth:2008rx}.  Even including all these bounds,
however, there is an allowed region when channeling or streams are
included, as illustrated in \figref{damadirect}.

Unfortunately, the low mass DAMA region is very difficult to realize
in standard weakly-interacting massive particle (WIMP) frameworks.  In
the minimal supersymmetric standard model (MSSM) with gaugino mass
unification, for example, the neutralino mass is constrained to be
above 46 GeV~\cite{Abdallah:2003xe}.  This may be evaded by relaxing
gaugino mass unification. The cross section is, however, a much more
robust problem.  Spin-independent scattering requires a chirality flip
on the quark line.  In supersymmetric models with minimal field
content and other well-known WIMP frameworks, $\sigmaSI$ is thus
highly suppressed by Yukawa couplings.  Neutralino cross sections as
high as $8 \times 10^{-5}~\pb$ are possible and may explain the DAMA
signal~\cite{Bottino:2003iu}, but more typically, $\sigmaSI$ falls
short of this value by many orders of magnitude.

\ssection{WIMPless Models} WIMPless dark matter provides a framework
in which dark matter candidates with a wide range of masses naturally
have the correct thermal relic density~\cite{Feng:2008ya}.  In
WIMPless models, the standard supersymmetric model with gauge-mediated
supersymmetry breaking is supplemented by a hidden sector, consisting
of particles with no standard model (SM) gauge interactions.  The
hidden sector contains the WIMPless dark matter particle, which has
mass $m_X$ at the hidden sector's supersymmetry breaking scale and
interacts through hidden sector gauge interactions with coupling
$g_X$.  Supersymmetry breaking in a single sector is transmitted
through gauge interactions to both the MSSM and the hidden sector.  As
a result,
\begin{equation}
\frac{m_X}{g_X^2} \sim \frac{m_W}{g_W^2} \ ,
\end{equation}
where $m_W \sim 100~\gev - 1~\tev$ and $g_W \simeq 0.65$ are the weak
mass scale and gauge coupling.  Because the thermal relic density of a
stable particle is
\begin{equation}
{\Omega} \propto {1\over \langle \sigma v \rangle} \sim
\frac{m^2}{g^4} \ ,
\end{equation}
$\Omega_X \sim \Omega_W$, the thermal relic density of a typical WIMP.
Since this is known to be approximately the observed dark matter
density, these hidden sector particles also have approximately the
observed dark matter density, preserving the key virtue of WIMPs.  At
the same time, WIMPless dark matter need not have weak-scale mass, and
so provides a promising scenario to explain the DAMA signal in the low
mass region with parameters $m_X \sim 2-10~\gev$ and $g_X \sim 0.1$.

\ssection{Direct Detection} Of course, a valid DAMA explanation also
requires the correct $\sigmaSI$.  WIMPless dark matter has no SM gauge
interactions, but may have non-gauge interactions with SM particles
without spoiling the motivations detailed
above~\cite{Feng:2008ya,Feng:2008mu}.  In fact, intersecting brane
models motivate connector particles $Y$, charged under both SM and
hidden sector gauge groups, to mediate such
interactions~\cite{Cvetic:2001nr}.  Consider the interactions
\begin{equation}
{\cal L} = \lambda_f X \bar{Y}_L f_L
+ \lambda_f X \bar{Y}_R f_R \ ,
\label{connector}
\end{equation}
where $X$ is a scalar WIMPless candidate, the connectors $Y_{L,R}$ are
chiral fermions, and $f_{L,R}$ are SM fermions. These terms mediate
spin-independent $X$-nucleus scattering via $X q \to Y \to X q$ with
cross section
\begin{eqnarray}
\sigmaSI &=& \frac{1}{4\pi}
\frac{m_N^2}{(m_N + m_X)^2} \times \nonumber \\
&& \left[ \sum_q \frac{\lambda_q^2}{m_Y - m_X}
\left[ Z B^p_q + (A-Z) B^n_q \right] \right]^2  ,
\end{eqnarray}
where we have assumed $X$ is not its own anti-particle, and $Z$ and
$A$ are the atomic number and mass of the target nucleus $N$.  For the
light quarks $q=u, d, s$, $B^{p,n}_q = \langle p,n | \bar{q} q | p,n
\rangle \equiv m_{p,n} f^{p,n}_q / m_q$.  For the heavy quarks,
nucleon scattering arises through gluon couplings induced by triangle
diagrams with the quarks in the loop.  These diagrams can be computed
simply from anomaly considerations~\cite{Shifman:1978zn}, and one
finds $B^{p,n}_q = (2/27) m_{p} f^{p,n}_g / m_q$ for $q=c,b,t$, where
$f^{p,n}_g = 1 - f^{p,n}_u - f^{p,n}_d - f^{p,n}_s$.  Reasonable
values for the hadronic parameters are $B^p_u = B^n_d \simeq 6$,
$B^p_d = B^n_u \simeq 4$, $B^{p,n}_s \simeq 1$, and $f^{p,n}_g \simeq
0.8$~\cite{Ellis:2001hv}.

The connectors $Y$ are similar to 4th generation quarks.  They get
mass from both SM and hidden gauge couplings, and so we expect $m_Y
\sim \max(m_W, m_X)$; given that we are interested in the DAMA signal
with $m_X < m_W$, this implies $m_Y \sim m_W$.  The Yukawa couplings
$\lambda_q$ are model-dependent.  If all are ${\cal O}(1)$, these
couplings would violate flavor bounds.  We will assume that only
$\lambda_b$ and $\lambda_t$ are significant.  These are the least
constrained experimentally, and it is reasonable to assume that the
others are Cabbibo-suppressed. Top quark contributions to $\sigmaSI$
are suppressed by $m_t$, and so with this assumption, $\sigmaSI$ is
dominated by the coupling to the $b$ quark.

The results for $\sigmaSI$ for $X$-proton scattering as a function of
$m_X$ are given in \figref{damadirect} for various values of
$\lambda_b$.  For $\lambda_b \sim 0.5$, $\sigmaSI$ is in the required
range to explain the DAMA signal.  The WIMPless model therefore
matches both the required mass and cross section without difficulty.
Note that $\sigmaSI$ is much larger than is typical for WIMPs.  The
problem of chirality flip suppression noted above is solved by
introducing a heavy fermion as an intermediate state.  This
possibility was noted previously for scalar dark matter in another
context~\cite{Boehm:2003hm}.  In WIMPless models, this general
solution arises naturally, with the ``4th generation quarks''
$Y_{L,R}$ playing the role of heavy fermions.  As with any other
proposal that targets the low mass DAMA-favored region, this
explanation will be tested by progress in direct detection from, for
example, future analyses of CDMS data and ultra-low threshold
experiments~\cite{Barbeau:2007qi}.

\ssection{Indirect Detection} We now focus on the DAMA-favored
parameter region without dark matter streams, where $m_X \agt
5~\gev$. The interactions of \eqref{connector} will also mediate $X
\bar X \to b\bar b $ via $t$-channel $Y$ exchange.  These quarks will
hadronize and decay into showers of photons, $e^+$, $e^-$, and
neutrinos, providing interesting signals for indirect dark matter
detection at a variety of experiments.

The annihilation cross section for $X \bar X \to b\bar b$ is
\begin{equation}
\label{indirect}
\sigmabb v = {\lambda_b^4 \over 4\pi}
{m_Y ^2 \over (m_X ^2 +m_Y ^2)^2} \left( 1-\frac{m_b^2}{m_X^2}\right)^{3\over 2}\ .
\end{equation}
This depends on several unknown parameters.  However, requiring that
this WIMPless dark matter fit the low mass DAMA region fixes $m_X \ll
m_N, m_Y$.  In this limit, both $\sigmaSI$ and $\sigmabb v$ depend on
$\lambda_b$ and $m_Y$ only through the combination $\lambda_b^2 /
m_Y$.  Requiring that $\sigmaSI$ fit the DAMA-favored value then fixes
$\sigmabb v \simeq 6 \times 10^{-26}~\cm^3~\s^{-1} \simeq 2\,{\rm
pb}$, completely determining the predicted signal at indirect
detection experiments in this model.  The cross-section for WIMPless
dark matter to annihilate to hidden sector particles via hidden gauge
interactions is of approximately the same order of magnitude, yet the
total annihilation cross-section cannot be too large if the WIMPless
candidate is to be a significant dark matter component.  But for
WIMPless dark matter (unlike neutralinos) the precise relation between
the total annihilation cross section and the relic density depends on
some model-dependent factors, such as the ratio of hidden and visible
sector temperatures and the number of relativistic degrees of freedom
in the hidden sector~\cite{Feng:2008mu}.  For reasonable choices of
these parameters, this WIMPless model can have a relic density that is
$10-100\%$ of the observed density of dark matter.

We focus now on photon detection prospects~\cite{Bergstrom:1997fj}.
The shape of the photon spectrum is determined by the $b \bar{b}$
annihilation channel; it is given in Ref.~\cite{Baltz:2006sv}, and
$E^2 d\Phi/dE$ peaks at $E \approx m_X/25$.  The normalization is
determined by $\sigmabb v$, the experiment's angular resolution, and
the halo profile.  We assume an angular resolution $\Delta \Omega =
10^{-3}~\sr$ and choose an NFW halo profile~\cite{Navarro:1995iw},
which matches both the local density of $0.3~\gev~\cm^{-3}$ and the
halo mass of $7 \times 10^{11} M_\odot$ within 100
kpc~\cite{Widrow:2005bt}.  These halo parameters may vary by factors
of 2 or more. This is a moderate profile, with $\rho \propto r^{-0.8}$
in the Galactic center and cuspiness parameter $\bar{J} = 12$.  We
note, however, that the photon spectrum normalization may be smaller,
with profiles with $\rho \propto r^{-0.4}$ yielding $\bar{J} \sim 1$.

\begin{figure}
\resizebox{3.25in}{!}{
\includegraphics{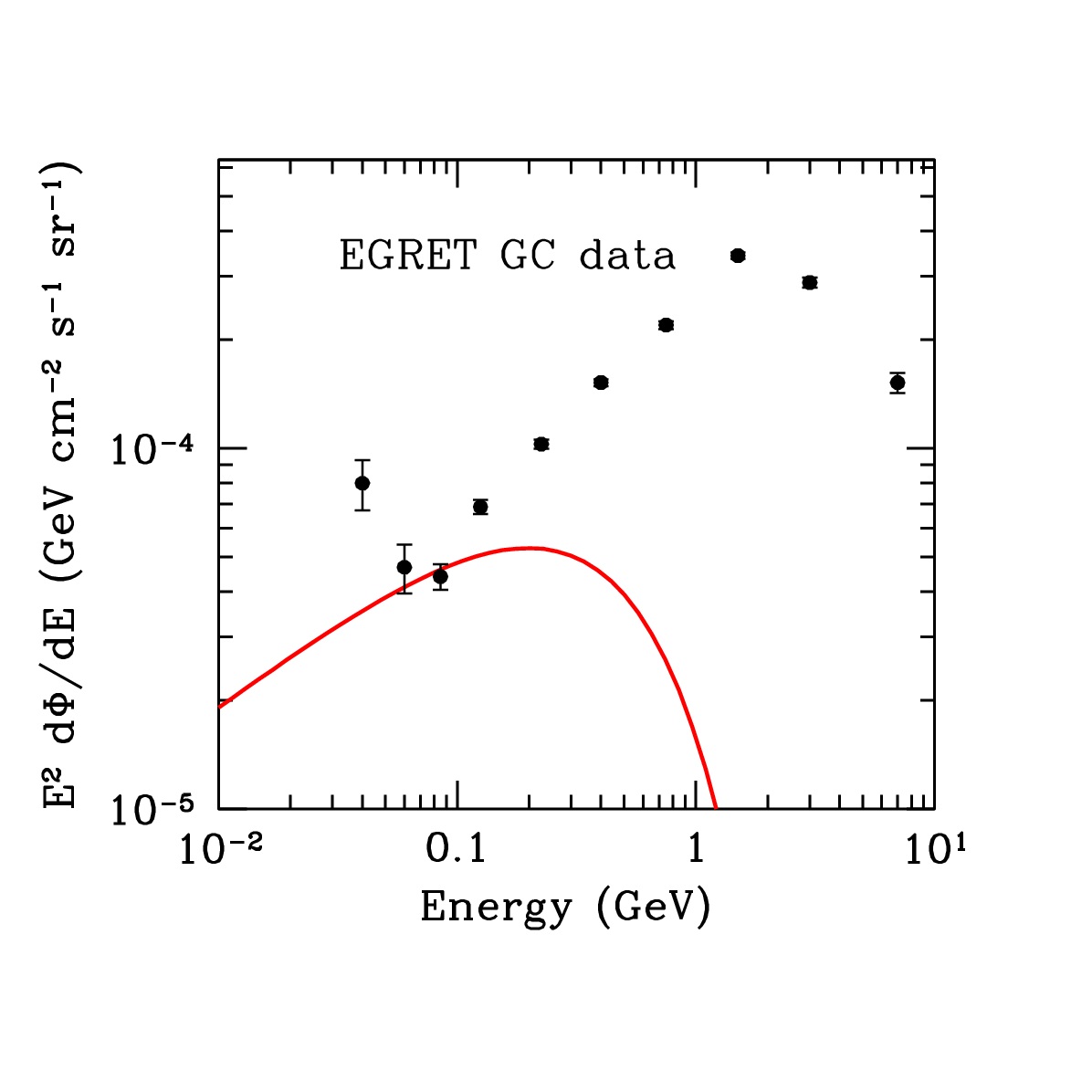}
}
\caption{Predicted $\gamma$-ray spectrum for a WIMPless model that
explains DAMA with $m_X = 6~\gev$ and $\sigmabb v = 6 \times
10^{-26}~\cm^3~\s^{-1}$ and an NFW halo profile (see text). The data
are EGRET's $\gamma$-ray spectrum from the Galactic
center~\cite{Hunter1997}.
\label{fig:gammaray}
}
\end{figure}

In \figref{gammaray} we plot the resulting $\gamma$-ray spectrum for
$m_X= 6~\gev$, and $\sigmabb v = 6 \times
10^{-26}~\cm^3~\s^{-1}$. Also plotted is EGRET's $\gamma$-ray spectrum
from the Galactic center region~\cite{Hunter1997}.  We see that, even
for a standard halo profile, WIMPless models predict spectra that may
be as large as allowed by current data.  Such signals provide a
promising target for the recently launched Gamma-ray Large Area Space
Telescope (GLAST).  As expected, the spectrum peaks near $0.2~\gev$.
Internal bremsstrahlung, so promising because its $E^2 d\Phi/dE$ peaks
very near $m_X$, is unfortunately suppressed in this case by the high
mass of the final state $b$ quarks~\cite{Bergstrom:1989jr}.
Nevertheless, the peak in $E^2 d\Phi/dE$ at $E \approx 0.2~\gev$
provides a specific observable prediction of this model.

Although we have focused on $\gamma$-rays from the Galactic center,
the large annihilation cross section may make it possible to detect
$\gamma$-rays from other sources in the halo, for example from the
diffuse emission away from the Galactic center or from dark
matter-dominated Galactic satellites. These latter objects may provide
a more robust signal given the reduced backgrounds and more
well-measured dark matter distributions.  For example, scaling the
results of Ref.~\cite{Strigari:2007at} to the masses and cross
sections for WIMPless models, we find fluxes integrated over solid
angles of $10^{-10}-10^{-9}~\cm^{-2}~\s^{-1}~\gev^{-1}$ from the
nearest satellites, also within reach of GLAST.

The annihilation process $X \bar X \to b \bar b$ also produces
positrons~\cite{Rudaz:1987ry}.  For the model we consider, where
$\sigmabb v \simeq 6 \times 10^{-26}~\cm^3~\s^{-1}$, with certain
assumptions about the local dark matter
density~\cite{Moskalenko:1999sb}, the $e^+$ flux can be competitive
with that observed by HEAT~\cite{Barwick1997}.  Though this signal is
affected by astrophysical uncertainty in $e^+$ propagation,
experiments like PAMELA may be sensitive to wide regions of WIMPless
parameter space, providing another interesting channel for study.

\ssection{Collider Signatures} This WIMPless model also has
distinctive signatures for the Large Hadron Collider (LHC).  The most
dramatic signatures are from production of the exotic connector quark
multiplets $Y_{L,R}$.  These get mass from electroweak symmetry
breaking and are constrained by direct searches at the Tevatron,
corrections to precision electroweak observables, and
perturbativity~\cite{Kribs:2007nz}.  These constraints require
$260~\gev \alt m_Y \alt 600~\gev$.  In this mass range, the process $p
p \to Y \bar{Y} \to X \bar{X} b \bar{b}$ should be observable at the
LHC, and the combination of this signature with typical
gauge-mediation signatures, for example, long-lived sleptons,
multi-lepton or prompt photon events, is distinctive.  Monojet and
single photon signals from $pp \to X \bar X (j, \gamma)$ are also
possible.  These were judged unpromising in conventional WIMP
models~\cite{Birkedal:2004xn}, but may be more interesting in the
WIMPless models, where thermal relic constraints are effectively
decoupled from observable signal strengths.

In addition, if the WIMPless explanation for the DAMA results holds,
there are many striking implications for Higgs physics, which have
been explored in detail in the related context of 4th generation
quarks~\cite{Kribs:2007nz}.  The $Y$ connectors raise the Higgs boson
mass far above the typical supersymmetric limit of $130~\gev$,
alleviating fine-tuning and making supersymmetry compatible with the
golden Higgs signal region at the LHC.  They also enhance $\sigma(gg
\to h)$ by an order of magnitude, and strengthen the first-order
electroweak phase transition, making electroweak baryogenesis
viable~\cite{Kribs:2007nz}.  Comprehensive detection strategies for
this WIMPless model will have an interesting interplay with Higgs
searches and other studies of new physics at the LHC.

\ssection{Summary} One of the barriers to a theoretical understanding
of the DAMA result is the difficulty in finding suitable candidates to
explain it.  In contrast to WIMPs, the WIMPless model proposed here
easily matches the low mass and extremely high cross sections of the
low mass DAMA region, while preserving the WIMP motivation of the
naturally correct thermal relic density.  This explanation implies
specific and promising signals for other direct detection experiments
and for indirect searches, such as GLAST, and the LHC.  We note that
the required DAMA cross sections are ``extremely high'' only when
viewed from the WIMP viewpoint and are quite naturally achieved by the
introduction of heavy colored fermions.  This is a rather
straightforward way to explain DAMA, and much of the discussion above
will hold more generally in any model that explains DAMA in this way.

\ssection{Acknowledgments} This work was supported by NSF grants
PHY--0239817, PHY--0314712, PHY--0551164 and PHY--0653656, NASA grant
NNG05GG44G, and the Alfred P.~Sloan Foundation.  We are grateful to
L.~Bergstrom, D.~Finkbeiner, P.~Gondolo, M.~Kaplinghat, G.~Kribs,
R.~Mohapatra, S.~Murgia, A.~Pierce, H.~Tu and H.~Yu for discussions
and thank the KITP, University of Michigan and University of New
Mexico for hospitality.



\end{document}